\documentclass[aps,prd,preprint,groupedaddress]{revtex4-1}

\usepackage{amssymb}
\usepackage{amsmath}
\usepackage{graphicx}
\usepackage{enumerate}
\usepackage{mathtools}

\begin{document}

\title{Neutrino oscillations in a trapping potential}

\author{Lucas Johns}
\email[]{ljohns@physics.ucsd.edu}
\affiliation{Department of Physics, University of California, San Diego, La Jolla, California 92093, USA}


\begin{abstract}
A number of derivations of the standard neutrino oscillation formula are known, each one providing its own unique insights. Common to all treatments is the assumption that neutrinos propagate freely between source and detector, as indeed they do in all experiments thus far conducted. Here we consider how neutrinos oscillate when, contrary to the usual set-up, they are bound in a potential well. The focus in particular is on nonrelativistic neutrinos with quasi-degenerate masses, for which oscillations in free space are described by the same formula, to lowest order, as relativistic neutrinos. Trapping these particles engenders corrections to their oscillation frequencies because the interference terms are between discrete energy levels rather than continuous spectra. Especially novel is the frequency shift that occurs due to the dependence of the energy levels on the mass of the neutrino: this part of the correction is nonvanishing even in the extremely nonrelativistic limit, reflecting the fact that the neutrino mass states have different zero-point energies in the well. Building an apparatus that can trap neutrinos is a futuristic prospect to say the least, but these calculations nonetheless shine a light on certain basic aspects of the flavor-oscillation phenomenon.
\end{abstract}


\maketitle

\section{Introduction}

Fundamentally, neutrino oscillations are an interference phenomenon that occurs when it is unclear from the kinematics of a process which mass state was produced or detected. In a typical experiment, measurements are made at various distances from a source and an oscillatory pattern is charted out, with frequency $\omega_{ij}$ for the interference of mass states $\nu_i$ and $\nu_j$. One heuristic way to derive the standard expression for this frequency is to suppose that the mass states propagate as plane waves of equal momenta. After a time $t$ the phase difference is then $\left( E_i - E_j \right) t$. Expanding $E_i = \sqrt{p^2 + m_i^2}$ in the relativistic limit and equating $t$ with the propagation length $L$, the phase difference becomes $\omega_{ij} t \cong \delta m^2_{ij} L / 2 p$. The same result is obtained if the mass states are instead taken to have equal energies.

In reality neutrinos do not propagate as plane waves of definite momentum or energy, and much work has been done to ground the result more rigorously. A number of different approaches have been adopted \cite{giunti1991, giunti1993, rich1993, blasone1995, kiers1998, blasone1999b, cardall2000, akhmedov2009, kayser2010, akhmedov2011, kobach2018, ciuffoli2019}, and though they offer different perspectives on how the interference pattern comes about, they all ultimately agree on it what it looks like. In all cases the relativistic expansion of $\sqrt{p_i^2 + m_i^2}$ appears at one point or another, as it ought to for particles propagating freely in vacuum. (Alterations to the dispersion relation due to coherent scattering in medium, such as when particles propagate through the Earth, are easily accommodated \cite{wolfenstein1978, notzold1988}, though the resulting phenomenology in very dense astrophysical environments is still being worked out, as in Refs.~\cite{duan2010, duan2015, johns2016, chakraborty2016, johns2017, wu2017, johns2018} and many others.) Hence, while the interference of $e^{i p_i x - i E_i t}$ plane waves is an unrealistic model of an oscillation experiment, it is nevertheless capturing something essentially correct about the physics.

Flavor oscillations can be pushed on conceptually by asking what happens when plane waves are \textit{not} the energy eigenstates appropriate to the problem. This would be the case if a neutrino, instead of propagating freely from source to detector, were trapped in a bound state of a potential well. No longer is the particle described by a wave packet enveloping a continuous distribution of momenta and energies, as it is in transit between production and detection, but rather a superposition of stationary states with discrete energy levels. Oscillations depend on energy differences---and energy differences depend on how energy is quantized in the well.

In fact the problem comes to resemble more closely the oscillations of neutrons into antineutrons \cite{kerbikov2004, phillips2016} or mirror neutrons \cite{kerbikov2008}, but with crucial differences with respect to each. When a neutron oscillates into an antineutron, it either escapes the trap (if the neutrons are bound magnetically) or annihilates on the walls (if the neutrons are bound materially). Similarly, when a neutron oscillates into a mirror neutron, it no longer feels the trap, be it a magnetic field or a bottle, and therefore escapes. In this paper we will zero in on the peculiarity of neutrino oscillation, which is that they are driven by a difference in mass as opposed to a difference in a quantum number like strangeness (for kaons) or baryon number (for neutrons). The significance of this distinction is that even if $\nu_i$ and $\nu_j$ experience the same potential, because they differ in mass their energy eigenstates will not coincide.

Needless to say, any proposal for trapping neutrinos in the real world would be a speculative one. We will not be concerned here with practicalities or feasibility, since our aim is merely to assess what effect there \textit{would} be, could such an experiment be carried out. In the backs of our minds, however, we might imagine a magnetic or material trap fashioned after the ones used for neutrons, setting aside the facts that neutrino magnetic moments are constrained to be very small \cite{studenikin2016} and that neutrino total reflection would be hindered by $G_F$. In analyzing neutrino oscillations in a harmonic potential we will assume for simplicity that the trap acts on the mass states, possibly with different strengths. If this were realized in a magnetic trap, it would correspond to the neutrino being a Dirac particle with a magnetic moment matrix that is diagonal in the mass basis.

Throughout this paper we will only be considering the trapping of neutrinos that are nonrelativistic. Oscillations occur in this limit provided that the mass states are quasi-degenerate: the uncertainty $\sigma_E$ in the energy must be large enough to encompass the mass splitting, which puts $\delta m_{ij}$ much below $m_i \approx m_j$ \cite{akhmedov2017}. Incidentally, there is an abundant source of neutrinos that are known to be at least partly nonrelativistic, namely the cosmic neutrino background (C$\nu$B). Since $| \delta m_\odot |$ and $| \delta m_\textrm{atm} |$ both exceed the current temperature of the C$\nu$B, at least two of the three neutrino mass states must be nonrelativistic today. If the lightest mass turns out to be comfortably above $\sim 0.1$ meV, then all three are nonrelativistic, and if it is at least several tens of meV, then quasi-degeneracy applies as well. Ideas for detecting the C$\nu$B mechanically, though not necessarily trapping them, go back a long way. Most proposals, but not all \cite{stodolsky1975}, relied on $G_F^2$ effects \cite{cabibbo1982, langacker1983}, rendering them unpromising. In any case, the C$\nu$B would not be ideal for the purpose of studying oscillations in a trap, since the individual mass states no longer overlap spatially. A viable source, if one can be found at all, would have to be found elsewhere.

We simply take it as our premise that a nonrelativistic neutrino has somehow found itself bound in a potential well. The mass states, which we assume to be quasi-degenerate, coexist in the trap, and the various stationary states associated with one mass interfere with the stationary states of the other masses and with the other stationary states of the same mass. All of these interference terms contribute to flavor oscillation. Perhaps most interesting is the effect on the oscillation frequency that occurs when the strength of the potential is itself dependent on $m_i$, as in a harmonic potential with a force constant $k$ that is \textit{in}dependent of $m_i$. Neutrino mass then enters the oscillation frequency in two separate ways: directly, through the difference $\delta m_{ij}$ in the rest-mass energies, and indirectly, through its effect on quantization.

In Sec.~\ref{secfree} a derivation of the standard transition probability using freely propagating wave packets is reviewed. This derivation, based closely on Ref.~\cite{akhmedov2009}, will be a helpful point of comparison for the calculations that follow. Since the probability $P_{ab}(L)$ for $\nu_a$ to transition to $\nu_b$ after a propagation length $L$ is not applicable to flavor oscillation in a trap, the autocorrelation function is introduced as an alternative and evaluated for neutrinos traveling in free space. In Sec.~\ref{sectrap} the general form of the autocorrelation function of a trapped neutrino is given and the interference phases are written out explicitly for the elementary cases of an infinite square well and a harmonic potential. These are compared to the free-space formula. In Sec.~\ref{secconc} we conclude.

\section{Oscillations in free space \label{secfree}}

In comparing oscillations in free space to oscillations in a trapping potential, it is essential that we study the same quantities in both cases. The item of interest in a neutrino oscillation experiment is the probability for a neutrino created in flavor state $\nu_a$ to be detected in flavor state $\nu_b$ a distance $L$ away. To establish a point of comparison, we begin by reviewing a derivation of this quantity. It is not straightforwardly adaptable to the trapping potential, however, at least so long as the neutrino remains in the well. We therefore also introduce another quantity in this section, the autocorrelation function, which will enable more direct comparison.

Following the lucid analysis of Akhmedov and Smirnov \cite{akhmedov2009} and working in $1+1$ dimensions for simplicity, a neutrino propagating in free space consists of wave packets, one for each $\nu_i$, which mix through the usual PMNS matrix $U$:
\begin{equation}
| \Psi_a ( x, t) \rangle = \sum_i U_{ai}^{*} \psi_i (x,t) | \nu_i \rangle. \label{bigpsi}
\end{equation}
Here the neutrino is taken to have been produced at $t=0$ in flavor eigenstate $\nu_a$. (Realistically, the emission time is not measured, and it is not critical to assume that it is. See Ref.~\cite{kobach2018} for a derivation that dispenses with definite emission time.) Each wave packet is a sum of plane waves:
\begin{equation}
\psi_i (x, t) = \int_{-\infty}^{\infty} \frac{dp}{\sqrt{2\pi}} ~f_i (p) e^{i p x - i E_i (p) t },
\end{equation}
where $f_i (p)$ is the wave-packet envelope and $E_i (p) = \sqrt{m_i^2 + p^2}$. Expanding around the average momentum $p_i$,
\begin{equation}
E_i (p) \cong E_i (p_i) + v_i ( p - p_i ),
\end{equation}
where
\begin{equation}
v_i = \frac{\partial E}{\partial p} \bigg|_{p_i}
\end{equation}
is the group velocity. Higher-order terms are dropped under the assumption that the wave-packet width $\sigma_p$ is small compared to the mean energy. Plugging in this expansion and shifting the integration variable,
\begin{equation}
\psi_i (x, t) \cong e^{i p_i x - i E_i (p_i) t} \int_{-\infty}^{\infty} \frac{dp}{\sqrt{2\pi}} ~f_i (p) e^{i p ( x - v_i t) }. \label{littlepsi}
\end{equation}
Hence, provided that it is sharply peaked, the $\nu_i$ wave packet retains its shape as it propagates, moving at velocity $v_i$ and developing the same overall phase as the plane wave of average momentum. For later convenience, we define
\begin{equation}
\tilde{\psi}_i (x- v_i t) = \int_{-\infty}^{\infty} \frac{dp}{\sqrt{2\pi}} ~f_i (p) e^{i p ( x - v_i t) }.
\end{equation}

When a measurement is performed, the propagating neutrino is projected onto the flavor eigenstate $\nu_b$. The wave packet of the detection state depends on the particular apparatus doing the measuring, but it has the general form
\begin{equation}
| \Psi_b^D (x - L) \rangle = \sum_i U_{bi}^{*} \psi_i^D (x-L) | \nu_i \rangle,
\end{equation} 
with the packet peaked at the location $L$ of the detector. Pulling out an overall phase factor, we rewrite
\begin{equation}
| \Psi_b^D (x - L) \rangle = \sum_i U_{bi}^{*} e^{i p_i (x - L)} \tilde{\psi}_i^D (x-L) | \nu_i \rangle,
\end{equation}
The transition amplitude for a measurement at time $t$ is then
\begin{align}
A_{ab} (L,t) &= \int_{-\infty}^{\infty} dx ~\langle \Psi_b^D (x-L) | \Psi_a (x,t) \rangle \notag \\
&= \sum_i U_{ai}^{*} U_{bi} e^{-i E_i (p_i) t + i p_i L} \int_{-\infty}^{\infty} dx ~\tilde{\psi}_i^{D*} (x-L) \tilde{\psi}_i (x - v_i t).
\end{align}
Note that the integral only depends on $t$ through the combination $L - v_i t$. We denote it as
\begin{equation}
G_i (L - v_i t ) = \int_{-\infty}^{\infty} dx ~\tilde{\psi}_i^{D*} (x-L) \tilde{\psi}_i (x - v_i t).
\end{equation}
The probability of a transition to $\nu_b$ occurring in the detector at any time is then
\begin{align}
P_{ab} (L) &= \int_{-\infty}^{\infty} ~ dt | A_{ab} (L, t) |^2 \notag \\
&= \sum_{i, j} U_{aj}^* U_{bj} U_{ai} U_{bi}^* \int_{-\infty}^{\infty} dt ~e^{i \delta\phi_{ij} (L, t)} G_i^*(L - v_i t) G_j (L - v_j t),
\end{align}
where the phase difference is
\begin{equation}
\delta \phi_{ij} = (E_i - E_j) t - (p_i - p_j)L.
\end{equation}
Expanding the momentum difference yields
\begin{align}
\delta p_{ij} &= \sqrt{E_i^2 - m_i ^2} - \sqrt{E_j^2 - m_j^2} \notag \\
&\cong \delta E_{ij} - \frac{\delta m^2_{ij}}{2 E},
\end{align}
where the subscript on the energy is dropped because the distinction between $i$ and $j$ is higher-order. Specializing to the relativistic limit, $v_{i,j} \cong 1$. Hence
\begin{equation}
\delta\phi_{ij} \cong - \delta E_{ij} (L - t) + \frac{\delta m^2_{ij}}{2 E} L.
\end{equation}
Shifting the integral,
\begin{equation}
P_{ab} (L) \cong \sum_{i, j} U_{aj}^* U_{bj} U_{ai} U_{bi}^* e^{i \omega_{ij} L} \int_{-\infty}^{\infty} dt' ~e^{i \delta E_{ij} t'} G_i^*(-t') G_j (-t'),
\end{equation}
where the oscillation frequency is the usual one:
\begin{equation}
\omega_{ij} = \frac{\delta m^2_{ij}}{2 E}.
\end{equation}
Observe that if $G_i$ and $G_j$ are very sharply peaked at $t' = 0$, which is the case if $\tilde{\psi}^D_i$ and $\tilde{\psi}^D_j$ are very sharply peaked at $x = L$, then all of the phase difference seems to come from interference between mass states with the same energy but different momenta, as argued heuristically in the introduction.

$P_{ab} (L)$ describes the measurements made in a typical experiment: flavor is seen to oscillate as a function of distance $L$ from the source. As noted at the beginning of this section, this quantity cannot be adapted to the study of neutrino oscillations in a trapping potential without making specific assumptions about the operation of the experiment. A more straightforward way to compare to oscillations of trapped neutrinos is to look at the autocorrelation function
\begin{equation}
\mathcal{A} (t) = \int_{-\infty}^{\infty} dx ~\langle \Psi_a (x, t) | \Psi_a (x, 0) \rangle.
\end{equation}
For the free-space neutrino described by Eqs. \eqref{bigpsi} and \eqref{littlepsi}, this is
\begin{equation}
\mathcal{A} (t) = \sum_i e^{iE_i (p_i) t} | U_{ai} |^2 \mathcal{G}_i(v_i t), \label{autofree}
\end{equation}
where
\begin{equation}
\mathcal{G}_i (v_i t) = \int_{-\infty}^{\infty} dx ~\tilde{\psi}_i^* (x - v_i t) \tilde{\psi}_i (x). \label{calg}
\end{equation}
With just two flavors and assuming $a = e$,
\begin{align}
| \mathcal{A} (t) |^2 &= \cos^4 \theta | \mathcal{G}_1 (v_1 t) |^2 + \sin^4 \theta | \mathcal{G}_2 (v_2 t) |^2 \notag \\
&~+ 2 \sin^2 \theta \cos^2 \theta ~\mathfrak{Re}\left[ \mathcal{G}_1^* (v_1 t) \mathcal{G}_2 ( v_2 t ) e^{i ( E_2 (p_2) - E_1 (p_1) ) t} \right]. \label{genfreeauto}
\end{align}
If the experiment is not sensitive to differences in the wave-packet shapes of the two mass states, then with $v_{1,2} \cong 1$, one obtains
\begin{equation}
| \mathcal{A} (t) |^2 \cong | \mathcal{G}(t) |^2 \left( 1 - \sin^2 2\theta \sin^2 \left( \frac{\delta m^2}{4 p} t + \frac{\delta p}{2} t\right) \right).
\end{equation}
Now the expansion is in $\delta E$, not $\delta p$, as it was for the transition probability. Note that since $\tilde{\psi}_i(x)$ only has considerable support over a width of $\sim \sigma_x$ around the origin, $|\mathcal{G} (t)|^2$ goes to 0 as $t$ increases, indicating that the wave packet at time $t \gtrsim \sigma_x / v$ no longer spatially overlaps the initial wave packet. This particular effect can be artificially negated by shifting 
\begin{equation}
\tilde{\psi}_i (x) \rightarrow \tilde{\psi}_i (x - v_i t)
\end{equation}
in Eq.~\eqref{calg}, leaving the conjugate untouched. Then the function becomes time-independent, and unitarity dictates that $| \mathcal{G} |^2 = 1$. The usual formula for the survival probability, up to the $\delta p$ term, is thereby recovered from the autocorrelation function. The  $\delta p$ term is unfamiliar because it is irrelevant to oscillation experiments; as per $P_{ab}(L)$, any direct contribution from the momentum or energy splitting is suppressed by the fact that $L \cong vt$.

Working now in the nonrelativistic limit, but still using $f_i (p) \cong f_j (p)$ and $v_{1,2} \cong v$, we find
\begin{equation}
| \mathcal{A} (t) |^2 \cong | \mathcal{G}(v t) |^2 \left( 1 - \sin^2 2\theta \sin^2 \left( \frac{E_2(p_2) - E_1(p_1)}{2} t \right) \right),
\end{equation}
where now
\begin{equation}
E_2(p_2) - E_1(p_1) \cong \delta m \left( 1 - \frac{p^2}{2 m^2} \right) + \delta p \frac{p}{m}. \label{nrfreediff}
\end{equation}
The mass difference $\delta m$ can itself be expanded in the mass-squared splitting,
\begin{equation}
\delta m \cong \frac{\delta m^2}{2 m_1} - \frac{( \delta m^2 )^2}{8 m_1^3}. 
\end{equation}
Keeping only the leading term in $\delta m^2$ and letting $T = p^2 / 2m$,
\begin{equation}
E_2(p_2) - E_1(p_1) \cong \frac{\delta m^2}{2 E} \left( 1 - \frac{T}{m} \right) + \delta p \frac{p}{m}, \label{nrfreediff2}
\end{equation}
We will find a formally similar result for trapped neutrinos, with the kinetic energy replaced by the bound-state energy. Note that the oscillation frequencies of relativistic and nonrelativistic neutrinos are dissimilar in that the first correction away from the relativistic limit gives
\begin{equation}
E_2(p_2) - E_1(p_1) \cong \frac{ \delta m^2}{2 p} + \delta p \left( 1 - \frac{m^2}{2 p^2} \right).
\end{equation}
That is, the correction shifts the part of the oscillation frequency proportional to $\delta p$ rather than the part proportional to $\delta m^2$.

\section{Oscillations in a trapping potential \label{sectrap}}

For simplicity, throughout this section we assume that the potential acts on the mass states, with no off-diagonal terms to facilitate transitions. We ignore subtleties like the question of how the neutrino got into the well in the first place, of how thoroughly it has lost its resemblance to its initial spatial profile (if, for instance, it tunneled in as a Gaussian wave packet with a width smaller than the size of the trap), and of what the time scales are on which wave-packet revival \cite{robinett2004} or tunneling out of the well occur. 

The trapped neutrino has wave function
\begin{equation}
| \Psi_a^T (x, t) \rangle = \sum_i U_{ai}^* \sum_n f_{i,n} \phi_{i,n} (x) e^{-i E_{i,n} t} | \nu_i \rangle,
\end{equation}
where $\phi_{i,n}(x)$ is the $n$th energy eigenstate associated with $\nu_i$ in the trap, having eigenvalue $E_{i,n}$ and coefficient $f_{i,n}$. To be explicit, the free-space and trapped wave functions are formally related by the replacements 
\begin{align}
E_{i,n} &\longleftrightarrow E_i(p_i) \notag \\
\sum_n f_{i,n} \phi_{i,n} (x) &\longleftrightarrow \int \frac{dp}{\sqrt{2\pi}} f_i (p) e^{ipx - i (p - p_i) v_i t}.
\end{align}
If $v_i = 0$, then $\phi_{i,n}(x)$ is simply replaced by $e^{ipx}$. We work with orthonormal $\phi_{i,n}$ and $\sum_n | f_{i,n} |^2 = 1$.

The Hamiltonian acting on $\nu_i$ is
\begin{equation}
\hat{H}_i = \hat{m} + \frac{\hat{p}^2}{2m_i} + \hat{V}_i(x),
\end{equation}
where $\hat{m}$ is the neutrino mass operator. We let $U_{i,n}$ denote the energy without the rest mass. That is,
\begin{equation}
E_{i,n} = m_i + U_{i,n}.
\end{equation}

One possible analogue of $P_{ab}(L)$ would be to place a detector in the trap, or vice versa, and calculate $\langle \Psi_b^D | \Psi_a^T (x,t) \rangle$ for some detector wave function. This procedure would reveal spatial dependence of the interference in a form that looks quite different from how it looks in free space. It would not, however, be as robust a measure as we might like, given its sensitive dependence on assumptions made about the detector. Recall, by comparison, that the only important fact assumed about the free-space detector of the previous section was that it localized the detected particle to the vicinity of $x = L$.

We focus instead on the autocorrelation function of $| \Psi_a^T \rangle$:
\begin{equation}
\mathcal{A} (t) = \cos^2 \theta \sum_n |f_{1,n} |^2 e^{i E_{1,n} t} + \sin^2 \theta \sum_n |f_{2,n} |^2 e^{i E_{2,n} t},
\end{equation}
hence
\begin{align}
| \mathcal{A} (t) |^2 &= \cos^4 \theta \sum_n \left[ |f_{1,n} |^4 + 2 \sum_{m > n} | f_{1,n} |^2 | f_{1,m} |^2 \cos \left( \left( E_{1,n} - E_{1,m} \right) t \right) \right] \notag \\
&~+ \sin^4 \theta \sum_n \left[ |f_{2,n} |^4 + 2 \sum_{m > n} | f_{2,n} |^2 | f_{2,m} |^2 \cos \left( \left( E_{2,n} - E_{2,m} \right) t \right) \right] \notag \\
&~+ 2 \sin^2 \theta \cos^2 \theta \sum_{n,m} | f_{1,m} |^2 | f_{2,n} |^2 \cos \left( \left( E_{2,n} - E_{1,m} \right) t \right). \label{gentrapauto}
\end{align}
Our interest here is in the time-dependence. Oscillation frequencies arise corresponding to interference between any two energy levels \textit{within} the spectrum of the same particle or \textit{across} the two spectra. Note the structural similarities to Eqs. \eqref{autofree} and \eqref{genfreeauto}, the general expressions for the autocorrelation function in free space: $e^{i E_i(p_i)t} \mathcal{G}_i (v_it)$ has been replaced by $\sum_n e^{i E_{i,n} t} | f_{i,n}|^2$. More interference terms appear in the trap, namely those across different energies of a single $\nu_i$, because $e^{i E_{i,n} t}$ has not been set to its average value and pulled out of the sum, as was done in free space.

A key difference is that in a potential well all time-dependence appears in sinusoidal form, reflecting bound-state interference. The dependence through $\mathcal{G}_i(t)$ in free space, on the other hand, reflects the time-dependence associated with the wave packet moving away from its initial position. The difference is more explicit if one considers $v_i = 0$ in free space, meaning that the wave packet is stationary on average, dispersing outward symmetrically. In that case time dependence drops out altogether. The cosines in the first two lines of Eq.~\eqref{gentrapauto} are signatures of bound states.

Just as we assumed the free-space $\psi_i (x,t)$ to be sharply peaked in momentum space about its average value $p_i$, we can assume here that $| f_{i,n} |$ is sharply peaked as a function of $n$ about the average $ \langle E_i \rangle$, with some spread $\sigma_E$. If $\sigma_E$ is small on the scale of the level spacings of $\hat{H}_i$ and $\hat{H}_j$, then there will only be one non-negligible $f_{i, n}$ for each of the masses. If $\sigma_E$ is also small on the scale of the mass splitting $\delta m$, then oscillations do not occur at all: this restriction is tantamount to identifying the neutrino as being in a definite mass state. Of course, it is also possible for $\sigma_E$ to be smaller than $\delta m$ but not smaller than the level spacings. That case reduces to the usual problem of a particle of fixed mass whose wave function is a superposition of multiple stationary states.

Let us now focus on the interference that occurs between the $\nu_1$ and $\nu_2$ states, since these are the analogue of neutrino oscillations in free space. First consider the infinite square well, with the center of the well at $x = a/2$. Then
\begin{equation}
E_{2,n'} - E_{1,n} = \delta m + U_{2,n'} - U_{1,n},
\end{equation}
where $U_{i,n} = n^2 \pi^2/ 2 m_i a^2$ as usual. Letting $n' = n + \delta n$,
\begin{equation}
E_{2,n'} - E_{1,n} \cong \frac{\delta m^2}{2 m} \left( 1 - \frac{U_{n}}{m} \right) + \delta n \frac{U_{n}}{n}. \label{squarediff}
\end{equation}
Note that there is a correction even in the limit that $\delta n$ goes to zero. This is because even when $\nu_i$ and $\nu_j$ are at the same level $n$, the energies are unequal due to the dependence of $U_{i,n}$ on $m_i$. Fixing $n$ and taking $a \rightarrow \infty$, the correction does disappear, echoing the vanishing of the correction in free space (Eq.~\eqref{nrfreediff2}) when the momentum goes to zero. But given finite $a$, the correction is nonvanishing as long as there is a particle in the trap. Quantization is thus apparent in the oscillation formula.

In the harmonic potential, with $k_2 = k_1 + \delta k$ and $n' = n + \delta n$ as before, we have
\begin{equation}
E_{2,n'} - E_{1,n} \cong \frac{\delta m^2}{2 m} \left( 1 - \frac{U_{n}}{2m} \right) + \delta n \sqrt{\frac{k}{m}} + \delta k \frac{U_{n}}{2k}, \label{harmonicdiff}
\end{equation}
with $U_{i,n} = \left( n + 1/2 \right) \sqrt{ k_1 / m_1 }$. As with the square well, there is a correction to the oscillation frequency even when the mass states occupy the same level in their respective potentials, and also when the parameter distinguishing the potentials, in this case $\delta k$, is set to zero. In fact, if the force constants are identical and the particle is in a superposition of the $\nu_1$ and $\nu_2$ ground states, the shift in the frequency relative to the standard expression is still nonvanishing:
\begin{equation}
E_{2,0} - E_{1,0} \cong \frac{\delta m^2}{2 m} \left( 1 - \frac{\sqrt{k / m}}{4 m} \right). 
\end{equation}
This persistent correction is due to the difference in the zero-point energies of $\nu_1$ and $\nu_2$ in the well.

The effects of trapping on neutrino oscillations become more pronounced as the average bound-state energy difference $\delta \langle U \rangle$ becomes comparable to the rest-mass energy difference $\delta m$. Such an arrangement is conceivable if the energy uncertainty $\sigma_E$ is large enough, but at the same time, in order to bring out the effects of the potential wall, $\sigma_E$ should not much exceed the level spacing. If the wave function is spread over a large number of energy levels, their discreteness will be obscured.

\section{Conclusion \label{secconc}}

All existing and proposed neutrino oscillation experiments---whether the source is a nuclear reactor, a supernova, or the cosmic surface of last scattering---consist, at the most basic level, of the same scheme. A neutrino is produced, propagates freely, and is detected. Another format is possible, however, if evolution in a potential well is substituted in for free propagation. The reason this possibility is not discussed is obvious: neutrinos are a challenge to detect, much less trap and manipulate.

Practical difficulties notwithstanding, the thought experiment of a neutrino oscillating in a potential well highlights the role of the energy eigenbasis appropriate to the problem. The discreteness of energy levels in a trap, and their dependence on neutrino mass, appear in the autocorrelation function of neutrino flavor (Eq.~\eqref{gentrapauto}), signaling a discrepancy with respect to the free-space formula (Eq.~\eqref{genfreeauto}). Differences are also evident in the individual oscillation frequencies contributing to the overall flavor development, as seen by juxtaposing Eq.~\eqref{nrfreediff2} for free space with Eqs. \eqref{squarediff} and \eqref{harmonicdiff} for an infinite square well and harmonic potential, respectively. Although the differences are not of observational importance, they are a fundamental aspect of flavor oscillations as a quantum phenomenon.

\begin{acknowledgments}
The author warmly thanks George Fuller for comments on the manuscript. This work was supported by NSF Grant No. PHY-1614864.
\end{acknowledgments}

\bibliography{all_papers}

\end{document}